# Multi-harmonic Quantum Dot Optomechanics in fused LiNbO$_3$-(Al)GaAs hybrids

Emeline D. S. Nysten[1,2], Yong Heng Huo[3,4], Hailong Yu[5], Guo Feng Song[5], Armando Rastelli[3], Hubert J. Krenner[1,2,*]

[1] Lehrstuhl für Experimentalphysik 1 and Augsburg Centre for Innovative Technologies (ACIT), Universität Augsburg, Universitätsstraße 1, 86159 Augsburg, Germany

[2] Nanosystems Initiative Munich (NIM), Schellingstraße 4, 80799 München, Germany

[3] Institute of Semiconductor and Solid State Physics, Johannes Kepler Universität Linz, Linz Institute of Technology, Altenbergerstraße 69, 4040 Linz, Austria

[4] Hefei National Laboratory for Physical Sciences at Microscale, University of Science and Technology of China, Shanghai branch, & CAS-Alibaba Quantum Computing Laboratory, Shanghai, 201315, China

[5] Nano-Optoelectronics Laboratory, Institute of Semiconductors, Chinese Academy of Sciences. Beijing, 100083, China

[*] corresponding author: hubert.krenner@physik.uni-augsburg.de

## Abstract

We fabricated an acousto-optic semiconductor hybrid device for strong optomechanical coupling of individual quantum emitters and a surface acoustic wave. Our device comprises a surface acoustic wave chip made from highly piezoelectric LiNbO$_3$ and a GaAs-based semiconductor membrane with an embedded layer of quantum dots. Employing multi-harmonic transducers, we generated sound waves on LiNbO$_3$ over a wide range of radio frequencies. We monitored their coupling to and propagation across the semiconductor membrane both in the electrical and optical domain. We demonstrate enhanced optomechanical tuning of the embedded quantum dots with increasing frequencies. This effect was verified by finite element modelling of our device geometry and attributed to an increased localization of the acoustic field within the semiconductor membrane. For moderately high acoustic frequencies, our simulations predict strong optomechanical coupling making our hybrid device ideally suited for applications in semiconductor based quantum acoustics.

## Abstract

## Introduction

Surface acoustic waves (SAW) have a longstanding tradition in the investigation and control of semiconductor nanostructures at radio frequencies (rf) [1–14]. Very recently, these propagating coherent phonons have been recognized as a versatile tool to control single quantum systems [15,16], quantum dots [17–20], nitrogen vacancy centers [21,22] or superconducting artificial atoms [23]. On one hand, the strain field of the SAW enable the dynamic tuning of the QD emission energy by deformation potential coupling [17,24]. On the other hand, its electric field allows carrier and spin transport and injection inside the QDs and control over their occupancy state [8,25–29]. SAWs can be generated directly on III-V semiconductors and derived heterostructures exploiting these materials' piezoelectric properties. Their low electromechanical coupling coefficient, (GaAs $K^2 \approx 0.07\%$), results in a moderately efficient conversion of the applied electrical power to acoustic power. The transfer of the QDs onto a strong piezoelectric substrate such as LiNbO$_3$, with $K^2 \approx 5\%$, could help enhance the coupling between a semiconductor heterostructure and the SAW field [4]. However, direct epitaxial growth of semiconductors on LiNbO$_3$ has so far only be achieved for two-dimensional transition metal dichalcogenides [14,30]. Thus, for most semiconductors, such hybrid devices can be fabricated by epitaxial lift-off and transfer of the QD membrane on a LiNbO$_3$ SAW-chip [31–37]. The latter is equipped with interdigital transducers (IDTs) in a delay line geometry which allow for efficient generation of SAWs. Because of the sensitivity of the SAW propagation to mass loading and surface defects, a study of the SAW propagation within the transferred membrane is crucial, before advancing such devices to more complex systems. Here we report such a study by analysing the propagation of the wave inside the membrane in the acousto-electric and acousto-optic domains using a SAW delay lines and individual QDs, respectively. Both effects are recorded in parallel as a function of the electrical frequency applied to the IDT. The advanced Split-52 geometry IDTs [24], enabled us to probe SAW propagation over a wide range of frequencies from $\approx 150$ MHz up to $\approx 600$ MHz. Over this wide range of frequencies, we observe a clear increase of the optomechanical coupling parameter for increasing frequency. This increase arises from an enhanced localization of the SAW's acoustic field inside the semiconductor. Our experimental findings are in good agreement with realistic finite element simulation of the hybrid structure.

## Sample design and methods

The sample is schematically described in Figure 1(a). It consists of a semiconductor membrane transferred by epitaxial lift-off (ELO) onto a LiNbO$_3$ SAW-chip. For the SAW-chip three Ti/Al (5 nm / 50 nm) interdigital transducers (IDTs) and a Pd adhesion layer (50nm) were defined on a 128° rotated Y-cut, X-propagating LiNbO$_3$ substrate prior to the transfer of the membrane. The SAW-chip consists of two delay lines. Such double-delay line layout sharing the center IDT is commonly used in SAW-based sensors [38,39]. For our purpose, we take advantage of its capability to directly compare the transmission of a SAW propagating on a free LiNbO$_3$ surface and that propagating across the (Al)GaAs-LiNbO$_3$ fused hybrid. The first delay line formed by IDT1 and IDT2 served as a free propagating reference. The second delay line formed by IDT2 and IDT3 contains the QD membrane. The chosen Split-52 geometry of the IDTs enabled the excitation of the fundamental SAW and of its 2$^{nd}$, 3$^{rd}$, and 4$^{th}$ harmonics $f_{SAW,n}$, $n = 1,2,3,4$ [24]. This tailored frequency characteristics allowed for the characterisation of the SAW propagation at different frequencies. The fundamental period of the IDTs was $\Lambda = 27.52\ \mu m$, setting $f_{SAW,1} = 145$ MHz for the Rayleigh-type SAW of phase velocity $c_{0,LiNbO_3} = 3990\ \frac{m}{s}$ propagating along the X-direction on the open LiNbO$_3$ surface.

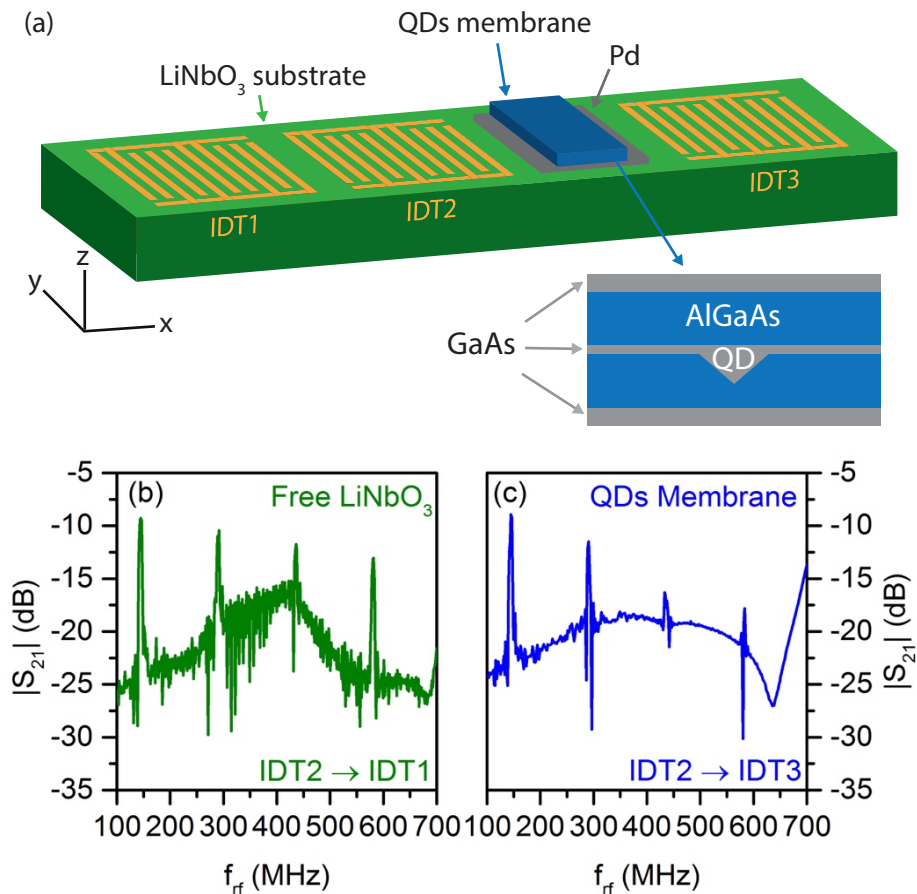

Figure 1: Schematic of the hybrid device composed of a QD membrane transferred by epitaxial lift-off on a LiNbO$_3$ SAW-chip patterned with two delay lines (a). A schematic of QD heterostructure is shown in the lower part. SAW transmission on free LiNbO$_3$ (IDT2 to IDT1) (b) and through the epilayer (IDT2 to IDT3) (c) measured with a vector network analyser at room temperature.

The 150 nm-thick active semiconductor membrane, depicted in Figure 1(a), was grown by molecular beam epitaxy on top of a 1μm-thick Al$_{0.75}$Ga$_{0.25}$As sacrificial layer. The membrane consists of 140 nm thick Al$_{0.33}$Ga$_{0.67}$As layer sandwiched between 5-nm-thick GaAs layers, which protect the AlGaAs material from oxidation. In the centre of the membrane, a ~~single~~ layer of GaAs QDs was fabricated by a droplet etching and filling technique [40]. ELO of the (Al)GaAs structure was performed by selectively etching the sacrificial layer in a dilute HF solution. The membrane was subsequently transferred onto the Pd layer. At the interface between the metal and the semiconductor a strong mechanical bond forms [41] which is indispensable for faithful transduction of the mechanical deformation into the semiconductor. From the as-transferred film, a rectangular-shaped part (with a lateral size of 450μm) was isolated by wet chemical etching [42] to ensure a constant interaction length between the SAW and the membrane and to avoid unwanted losses by scattering and reflections.

First, the fabricated samples were pre-characterized at room temperature by a vector network analyser (VNA) for both delay lines. Second, the SAW propagation inside the AlGaAs membrane was measured electrically via the SAW transmission in a delay line and optomechanically (at a temperature of 10 K) via the deformation potential coupling to individual QDs. Both effects were recorded in parallel as a function of the applied electrical radio frequency to IDT2: we applied rf pulses of variable frequency ($f_{rf}$) at a constant power $P_{rf} = 28$ dBm. A repetition rate of 50 kHz and a ON/OFF duty cycle of 1:19 were optimized to supress spurious heating of the sample. The rf signal transmitted through the membrane was detected at IDT3 in the time domain by a digital oscilloscope. The transmitted power by the SAW was quantified by a fast Fourier transform analysis. We determined the time-averaged optomechanical coupling of the QD using optical spectroscopy. All optical experiments were performed in a liquid helium flow cryostat in a conventional micro-photoluminescence (μ-PL) setup. Electrons and holes were photogenerated by an externally triggered pulsed diode laser at a wavelength of 660 nm and focused on the sample by a 50x microscope objective to a diffraction limited spot of diameter ~1.5 μm. The surface density of these QDs was < 1 μm$^2$, which allows to probe individual QDs and measure their optomechanical coupling to the SAW. The excitonic emission of individual QDs was then collected by the same objective, dispersed by a 0.5 m grating monochromator and detected by a liquid N$_2$ cooled silicon charge-coupled device (CCD). Because the laser pulse repetition rate was set to $f_{laser} \neq m \cdot f_{rf}$, the measured spectral broadening provides a direct measure of the local hydrostatic pressure at the position of the QD. The results of our experiments were complemented with and compared to finite element modelling (FEM) using Comsol Multiphysics. FEM was performed in a two-dimensional unit cell in which the structure of the epilayer was reproduced. Electrical and mechanical periodic boundary conditions were applied on the edges of the cell and the bottom boundary was maintained fixed. The Pd layer was treated as a non-piezoelectric elastic material with floating potential boundary conditions above and below. As the size of the unit cell was swept, the eigenfrequencies of the system were computed. By selecting the eigenfrequency corresponding to the SAW, the phase velocity of the SAW was computed from the size of the unit cell equal to the wavelength of the SAW. The local hydrostatic pressure at the position of the QD was also extracted directly from the simulation.

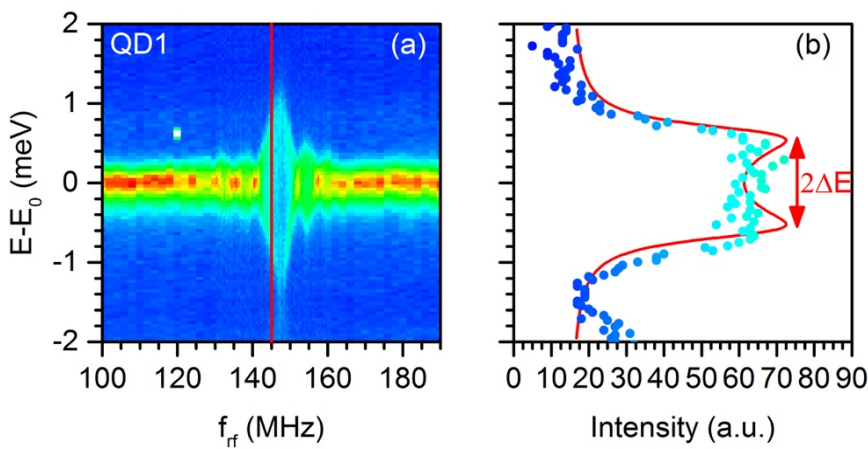

*Figure 2: (a) False colour plot of a time-integrated emission line of an individual QD as a function of the driving frequency of the IDT. (b) Fit of the SAW modulation of the QD emission line for one of the time-integrated spectrum (marked with a red line in (a)).*

## Results and Discussion

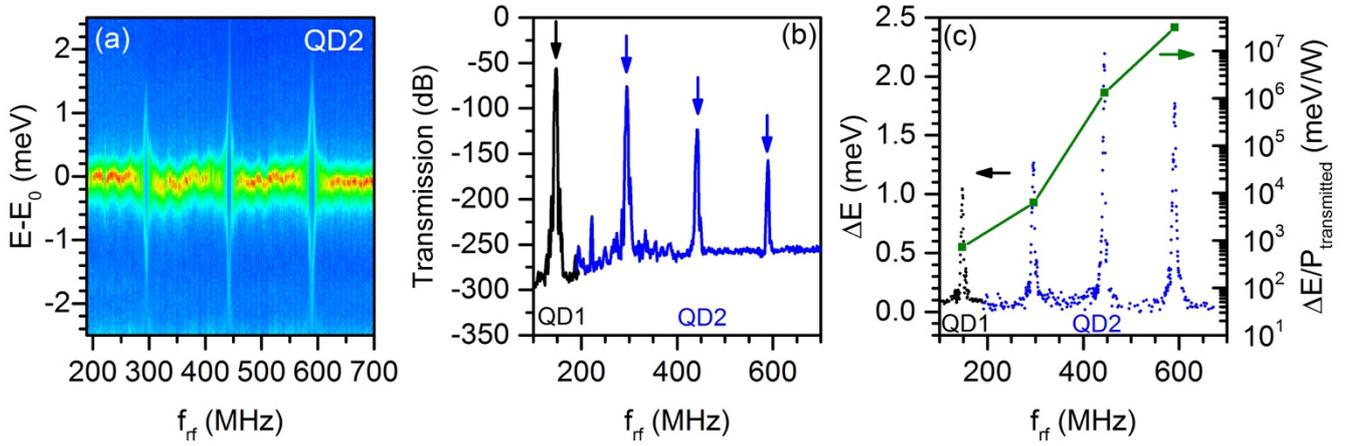

Figure 3: False colour plot of a time-integrated individual QD emission line as a function of the driving frequency of the IDT for the 2$^{nd}$, 3$^{rd}$ and 4$^{th}$ harmonics of a Spit-52 IDT. (b) SAW power transmission through the QD membrane as a function of the driving frequency. (c) Spectral modulation ($\Delta E$) due to the deformation potential coupling of the QD 1 and 2 extracted through the fit of the spectrum and its peak value divided by the power transmission at that point as a function of the driving frequency.

Figure 1(b) and (c) present the magnitude of the SAW transmission on free LiNbO$_3$ (IDT2 to IDT1) and across the QD membrane (IDT2 to IDT3) at room temperature. In both cases, the expected fundamental SAW is detected at $f_{SAW,1} = 145$ MHz, and its harmonics at $f_{SAW,2} = 292$ MHz, $f_{SAW,3} = 436$ MHz and $f_{SAW,4} = 582$ MHz. A closer comparison of the two delay lines shows that these frequencies match for both delay lines, even though the membrane is expected to act as an additional dispersive acoustic overlayer, which alters the sound wave's phase velocity. In fact, direct comparison between the data recorded from the reference and hybridized delay lines in Figure 1 (b) and (c), respectively, shows that the transmission of the higher harmonics $m = 3, 4$ is significantly reduced by the presence of this membrane. This observation could be intuitively explained by additional mechanical losses of the wave induced by the membrane. As the frequency increases or, equivalently, the wavelength decreases, the SAW deformation is localised closer to the surface and is more sensitive to scattering at edges and attenuation within the epilayer.

To quantify the coupling of the sound wave to the membrane, we performed a full optomechanical characterisation by detecting the spectral response of the emission of individual QDs as a function of $f_{rf}$ applied to IDT2. As this SAW couples to the membrane, the embedded QD is dynamically strained. Its sharp emission line is modulated by the deformation potential coupling [17,24,28]. In our time-integrated spectrum, this modulation leads to the observation of a broadening of the emission line [17]. For small SAW amplitudes as used in our experiments, the observed broadening scales linear with the amplitude [27,28]. Thus, the observed spectral broadening of the QD emission lines provides a direct measure of the SAW amplitude [42–44]. We note, that this broadening in fact consists of a series of acoustic sidebands split by precisely the SAW phonon energy. These sidebands have been observed by high-resolution spectroscopy for InAs/GaAs QDs [20,45] and NV centers in diamond [21]. The result of such an experiment is presented in Figure 2 (a) for QD1 in the frequency range close to the IDT's fundamental resonance. The emission intensity is plotted in false-color representation as a function of photon energy ($E$) and $f_{rf}$. As the IDT is driven at a frequency far from its resonant frequency, the input power is reflected by the transducer and no SAW is excited on the sample, thus the emission line of the QD remains unchanged. As the driving frequency approaches $f_{SAW,1}$, a larger fraction of the applied electrical power is transferred to the SAW. This broadening reaches a maximum at $f_{rf} = f_{SAW,1} = 148$ MHz, the fundamental SAW resonance of the IDT. Note that this resonance is slightly shifted to higher frequencies compared to that in Figure 1 due to an increase of the SAW's phase velocity at low temperatures. To quantify the measured optomechanical response, we fit the data using

$$I(E) = I_0 + f_{rf} \frac{2A}{\pi} \int_0^{1/f_{rf}} \frac{w}{4*\left(E-\left(E_0+\Delta E*\sin(2\pi \cdot f_{rf} \cdot t)\right)\right)^2 + w^2} dt.$$

*Equation 1*

This function corresponds to a time-integrated spectrum of a Lorentzian line of constant width ($w$) and intensity ($A$), whose centre wavelength ($E_0$) is sinusoidally modulated in time with amplitude $\Delta E$. Figure 2(b) shows an example of a best fit of Equation 1 (line) to our experimental data (symbol) at $f_{rf} = 145$ MHz, marked by the red line in panel (a).

We applied this method to investigate the coupling of the SAW to the semiconductor membrane. Figure 3 (a) shows an exciton emission line of a second QD, QD2, recorded over a wide range of frequencies, 190 MHz $\leq f_{rf} \leq$ 700 MHz. As $f_{rf}$ is tuned, we observe a pronounced broadening of the QD emission line for the three overtones of the fundamental SAW at $f_{SAW,2} = 297$ MHz, $f_{SAW,3} = 444.5$ MHz and $f_{SAW,4} = 591$ MHz.

In Fig. 3 (b) we plot the electrically detected transmitted SAW signal as a function of $f_{rf}$. The data plotted as black and blue lines were recorded simultaneously with the optical data from QD1 and QD2, respectively. The observed transmission peaks faithfully reproduce the maximum dynamic spectral broadening of the QD emission lines in the optical data marked by arrows. Moreover, the amplitude of the transmitted SAW intensity monotonously decreases with increasing frequency, *i.e.* decreasing SAW wavelength. This reduction is in fact expected due to increased scattering losses at the edges of the semiconductor membrane. The Rayleigh SAW is confined to approximately one acoustic wavelength. Thus, shorter wavelength SAWs are more susceptible to discontinuities such as the edge of our ~150 nm thick semiconductor membrane than long wavelength (low frequency) SAWs.

Interestingly, this clear trend is not observed for the spectral modulation of the QD emission line. This is seen in Figure 3 (b), where we plot the spectral broadening, $\Delta E$, extracted from the data of QD1 (black symbols) and QD2 (blue symbols) as a function of $f_{rf}$. In this data, $\Delta E$ shows precisely the opposite trend, i.e., it increases as $f_{rf}$ is increased up to $f_{SAW,3}$. Only for the highest SAW resonance at $f_{SAW,4}$ a moderate reduction is observed.

Next, we analysed $\Delta E/P_{transmitted}$, the observed spectral broadening referenced to the transmitted acoustic signal (in units of W). Thus, we account for the reduction of the acoustic power detected at IDT3 for higher $f_{SAW}$. The obtained $\Delta E/P_{transmitted}$, plotted in green in the same plot in Figure 3 (c) on a logarithmic scale (right axis), indicates a dramatic increase of the coupling efficiency of SAWs to the QD with increasing frequency.

To quantify the enhanced sound-matter interactions and the coupling of QDs to the acoustic field in our hybrids, we performed established finite element method (FEM) calculations [27]. In particular, we extracted the derivative of the hydrostatic pressure at the position of the QDs on the vertical displacement at the surface of the semiconductor membrane $\frac{\partial p(@QD)}{\partial u_z(z=0)}$ and the corresponding optomechanical coupling parameter $\gamma_{om} = \frac{\partial(\Delta E_{QD})}{\partial u_z}\bigg|_{z=0}$ [24,46]. In Figure 4 (a-f) we present calculated profiles of both quantities within a one acoustic wavelength wide unit cell. The assumed geometry is shown schematically as an inset of Figure 4 (g). In the simulation data shown in Figure 4 (a-c), we assumed a 150 nm membrane as used in the studied samples and calculated the hydrostatic pressure profile for $f_{SAW} = 103$ MHz, (a) 500 MHz (b) and 1218 MHz (c). Our simulations confirm an enhanced localization of the acoustic wave within the semiconductor membrane for higher frequencies, confirming the observed increase of the optomechanical response of the QDs. This effect becomes even clearer in the data shown in Figure 4 (d-f). For these simulation, we assumed $f_{SAW} \approx 2500$ MHz and varied the membrane thicknesses 150 nm, 300 nm and 450 nm. The calculated profiles clearly confirm the increased localization of the acoustic field within the semiconductor for thicker membranes. This effect leads to guiding of the acoustic wave inside the slow ($c_{0,GaAs} = 2900 \frac{m}{s}$) medium.

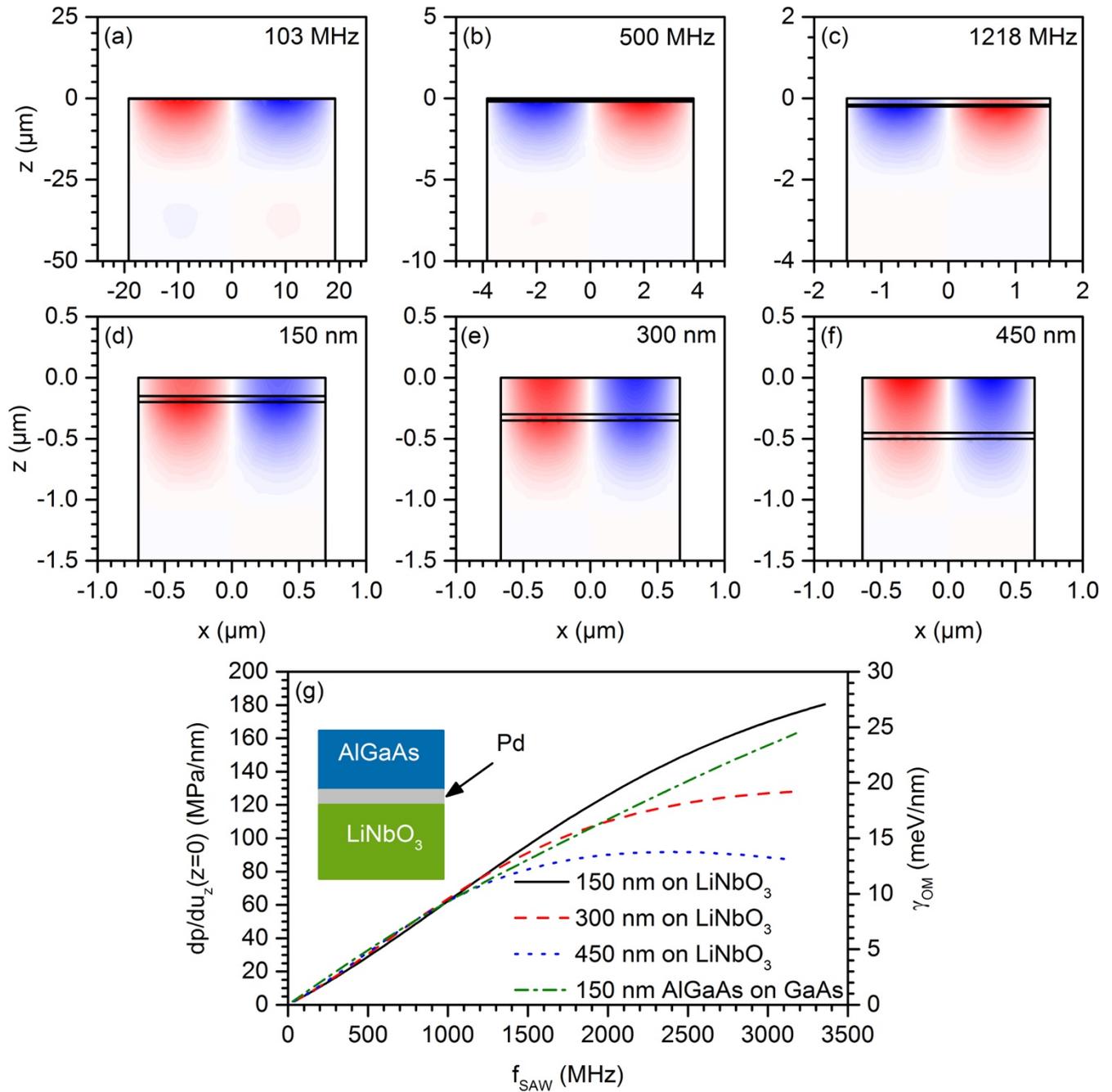

Figure 4: (a), (b) and (c) Colour plots of the pressure per nm vertical displacement (see text) resulting from the SAW inside a unit cell with a membrane thickness of 150 nm for a SAW frequency of 103 MHz, 500 MHz and 1218 MHz with a range of [-11, 11 MPa/nm], [-56, 56 MPa/nm] and [-138, 138 MPa/nm] respectively. (d), (e) and (f) The membrane thickness is varied between 150, 300 and 450 nm and the frequency is kept constant ≈2500MHz. The pressure per nm vertical displacement ranges from (d) [-266, 266 MPa/nm], (e) [-175,175 MPa/nm] and (f) [-148, 148 MPa/nm]. (g) Pressure per nm of vertical displacement at the position of the QDs and optomechanical coupling parameter as a function of the frequency of the SAW for three different membrane thicknesses and for the QD sample before ELO transfer onto LiNbO₃. The inset shows the geometry used for the finite element simulation.

To study this waveguiding effect, performed such calculations over a wide range of $f_{SAW}$. The obtained data is plotted in Figure 4 (g) for three membrane thicknesses 150 nm (black solid line), 300 nm (red dashed line) and 450 nm (blue short dashed line). For low frequencies $f_{SAW} \lesssim 1$ GHz, both $\frac{\partial p}{\partial u_z}$ and $\gamma_{om}$ increase linearly with $f_{SAW}$ for all three membrane thicknesses. In this regime, the thickness of the membrane is small compared to the acoustic wavelength, the lengthscale over which the acoustic fields are confined. Thus, the LiNbO₃ substrates directly induces the hydrostatic pressure within the semiconductor membrane and the derived linear dependences of $\frac{\partial p}{\partial u_z}$ and $\gamma_{om}$ on $f_{SAW}$ mirror the linear increase of $\frac{\partial p}{\partial u_z}$ at the interface between LiNbO$_3$ and the Pd/semiconductor stack. The calculated values for $\gamma_{om}$ for our LiNbO$_3$-(Al)GaAs hybrids are in the same range as the one achieved in our previous work [24] for a SAW directly excited on a similar (Al)GaAs heterostructure. For our hybrid, we obtain $\gamma_{om} = 1.45 \frac{\text{meV}}{\text{nm}}$ at $f_{SAW} = 180$ MHz and $\gamma_{om} = 1.8 \frac{\text{meV}}{\text{nm}}$ for the sinusoidal SAW at $f_{SAW} = 182.7$ MHz we reported in [24]. At high frequencies, $f_{SAW} > 1$ GHz, the linear dependencies of $\frac{\partial p}{\partial u_z}$ and $\gamma_{om}$ on $f_{SAW}$ break down and both quantities saturate. This saturation marks the transition from the LiNbO$_3$-dominated to the acoustic waveguiding regime. In this regime, the acoustic field shifts from the material of high phase velocity (LiNbO$_3$) into the material of lower phase velocity to finally becomes fully localized in the slow medium. In the latter, the acoustic fields become tightly confined within the semiconductor membrane. This confinement leads to a pronounced dependence of the optomechanical coupling parameter $\gamma_{om}$ on the membrane thickness. For thin membranes, the acoustic fields are confined within a smaller lengthscale compared to thick membranes. This in turn gives rise to a larger saturation value for thin membranes. The waveguiding effect in the hybrid LiNbO$_3$-(Al)GaAs device also increase the mechanical coupling of the dot to the SAW in comparison to the direct excitation of the SAW on the heterostructure as presented in figure 4(g). For our hybrid LiNbO$_3$-(Al)GaAs geometry, FEM simulations predict $\gamma_{om} > 25 \frac{\text{meV}}{\text{nm}}$ for $f_{SAW} \approx 3$ GHz, which is larger than the value predicted for a direct excitation on the QD sample and the previously studied monolithic device.

## Conclusions

In summary, we realized fused LiNbO$_3$-(Al)GaAs hybrid devices with enhanced acousto-optic coupling. We show that a SAW launched on the strongly piezoelectric LiNbO$_3$ substrate couples into the semiconductor membrane, propagates and couples back to the LiNbO$_3$ substrate on which it can be detected electrically. In parallel we monitored the interaction between individual QDs within the semiconductor heterostructure and the dynamic pressure field of the propagating SAW. In this complementary experiment, we confirm the strong optomechanical coupling precisely at frequencies a SAW is generated on the host substrate. From FEM modelling of our structure we derive optomechancial coupling parameters $\gamma_{om} > 25 \frac{\text{meV}}{\text{nm}}$ for thin membranes and moderately high SAW frequencies $f_{SAW} \approx 3$ GHz. Such large optomechanical coupling parameter arises from acoustic waveguiding and strong localization of the acoustic field within the material of slower phase velocity, the semiconductor membrane. In turn, this effect makes our realized hybrid device ideally suited for quantum acoustic devices based on high quality SAW resonators [16,47] for coherent frequency transduction between the nanomechanical and optical domain [48]. Our approach could be readily extended to alternative material combination such as AlN/diamond [21,49].

## Acknowledgements

This project has received funding from the European Union's Horizon 2020 research and innovation programme under the Marie Sklodowska-Curie grant agreement No 642688 (SAWtrain) and the grant no. 645776 (ALMA), the Deutsche Forschungsgemeinschaft via the Emmy Noether Program (KR3790/2-1) and the Cluster of Excellence "Nanosystems Initiative Munich" (NIM). H. J. K. and E. D. S. N. thank Achim Wixforth for his continuous support and inspiring discussions.